\documentclass[11pt]{article}

\usepackage[usenames]{color}
\usepackage{amssymb}
\usepackage{amsmath}
\newcommand{\half}{\frac{1}{2}}
\newcommand{\del}{\partial}
\newcommand{\ep}{\epsilon}
\newcommand{\bra}[1]{\langle #1 |}

\newcommand{\bracket}[2]{\langle #1| #2 \rangle}

\newcommand{\nn}{\nonumber}
\newcommand{\Pmatrix}[1]{\begin{pmatrix} #1 \end{pmatrix}}

\newcommand{\bea}{\begin{eqnarray}}
\newcommand{\eea}{\end{eqnarray}}

\newcommand{\ap}{\alpha}
\newcommand{\bt}{\beta}
\newcommand{\epp}{\sqrt{2\ep}}
\newcommand{\sqr}{\sqrt{\ap\bar\ap}}

\begin{document}

\begin{titlepage}
\title{
\hfill\parbox{4cm}
{\normalsize KUNS-1938\\{\tt hep-th/0410197}}\\
\vspace{1cm}
{\bf Massless Boundary Sine-Gordon Model\\
 Coupled to External Fields}
}
\author{
Hisashi {\sc Kogetsu}
\thanks{{\tt kogetsu@gauge.scphys.kyoto-u.ac.jp}}
\\[5pt]
{\small \it Department of Physics, Kyoto University, Kyoto 606-8502, Japan}
\\[5pt]
\quad and \quad\\[5pt]
Shunsuke {\sc Teraguchi}
\thanks{{\tt teraguch@phys.ntu.edu.tw}}
\\[5pt]
{\small \it Department of Physics, National Taiwan University, Taipei 106,
Taiwan}\\
{\small \it National Center for Theoretical Sciences at Taipei, National Taiwan
University, Taipei 106, Taiwan}
}
\date{\normalsize October, 2004}
\maketitle
\thispagestyle{empty}

\begin{abstract}
We investigate a generalization of the massless boundary sine-Gordon model with
conformal invariance,
which has been used to describe an array of D-branes (or rolling tachyon).
We consider a similar action whose couplings are replaced with external fields
depending on the boundary coordinate.
Even in the presence of the external fields, this model is still solvable,
though it does not maintain the whole conformal symmetry.
We obtain, to all orders in perturbation theory in terms of the external
fields, a simpler expression of the boundary state
and the disc partition function.
As a by-product, we fix the relation between the bare couplings and the
renormalized couplings
which has been appeared in papers on tachyon lump and rolling tachyon.
\end{abstract}

\end{titlepage}

\section{Introduction}

Investigations of two dimensional quantum field theories play
significant roles in several fields of physics.
Especially, solvable models with boundary interactions
find various applications
in condensed matter physics (for example, \cite{condensed}), string theory and
purely theoretical interests.
Boundary sine-Gordon model \cite{Ghoshal:1993tm} is one of such models.

Recent years, in the field of string theory, the massless boundary sine-Gordon
model with conformal invariance
has been often used to represent an array of localized D-branes
\cite{Sen:1999mh}
or a time-dependent
solution of string theory called rolling tachyon \cite{Sen:2002nu,Sen:2002in}.
(For a recent review of these topics, see \cite{SenReview}.)
The action of this model is given by\footnote
{In this paper, we take closed string picture, $0\leq\sigma\leq 2\pi$. The
boundary lies at $\tau=0$, where we conventionally describe the coordinate
$\sigma$ as $\theta$. And we use the convention of $\alpha^\prime=2$.}
\bea
S=\frac{1}{8\pi}\int d\sigma d\tau \del_a X\del^a X -\half \int d\theta
\left(ge^{\frac{iX(\theta)}{\sqrt{2}}}+\bar
ge^{\frac{-iX(\theta)}{\sqrt{2}}}\right). \label{originalBSG}
\eea
(For the case of rolling tachyon, the Wick-rotated version of this action is
used. In this paper, we shall concentrate on the above one.)
This model has been investigated in the papers
\cite{Callan:1993mw,Callan:1994ub,Polchinski:1994my,Recknagel:1998ih} in detail
and proved to have conformal invariance.
The boundary state for the boundary condition derived from this action
was calculated in \cite{Callan:1994ub},
\bea
\bra{B}=\bra{N}e^{-i\pi(g_rJ^+_0+\bar g_rJ^-_0)},\label{originalBS}
\eea
where $\bra{N}$ is the Neumann boundary state in the free theory at the
self-dual radius, $R=\sqrt{2}$.
$J^+_0$ and $J^-_0$ are zero-modes of holomorphic $SU(2)$ currents.
$g_r$ and $\bar g_r$ are some renormalized couplings which should be determined
by the original couplings, $g$ and $\bar g$, appearing in the action
(\ref{originalBSG}).
It is well-known that the absolute value of the renormalized couplings are in
the range of $ [-\half,\half] $.
However, the explicit relation between these two kinds of couplings has not
been given except for the case of some special limits.

In this paper, we consider a model with the following generalized boundary
interaction,
\bea
S_{\rm int}= - \half\int
d\theta\left[g(\theta)\exp{\left(\frac{iX(\theta)}{\sqrt{2}}\right)}+\bar{g}(\theta)\exp{\left(\frac{-iX(\theta)}{\sqrt{2}}\right)}\right],
\label{interaction}
\eea
where the external field $g(\theta)$ is an arbitrary function whose period is
$2\pi$ and
$\bar g(\theta)$ is the complex conjugate of $g(\theta)$.
Due to the explicit coordinate dependence of the external field, this model
does not have conformal symmetry.
Nevertheless, we can define this theory as a conformal perturbation theory and
solve it to all orders in perturbation theory.
We calculate the boundary state for this theory with the help of differential
equations.
In this process, we can also fix the relation between the bare couplings and
the renormalized ones
in the original theory (\ref{originalBS}), mentioned above.
Because the boundary state possesses the all information about the boundary
interaction,
we can easily calculate other quantities of this theory.

The outline of this paper is as follows. In section 2, we
perturbatively evaluate the
boundary state for the massless boundary sine-Gordon model with arbitrary
external fields. In this calculation we
shall fix our regularization scheme and notations.
In section 3, we derive the boundary state to all orders in
perturbation theory by introducing another
technique, namely, solving differential equations.
In section 4, we calculate the disc partition function and the correlation
functions of this model, using the results of section 3, and mention a symmetry
of this model.
In section 5, we summarize and discuss our results in this paper.

\section{Regularization and Some Perturbative Calculations}

In this section, we shall apply the technique used in
\cite{Callan:1994ub} to our model (\ref{interaction})
and obtain some perturbative results for its boundary state.
Though the application is straightforward, we shall provide this section for
fixing our conventions and regularization scheme.

In order to make the following calculations well-defined, we slightly shift the
positions of the boundary interactions to the inside of the disc by a short
distance $\ep$,
\bea
:\exp{\left(\frac{\pm iX(\theta,-\ep)}{\sqrt{2}}\right)}:, \label{regint}
\eea
This distance $\ep$ is the parameter for our regularization.
Note that, in order to define the above operators, we have used a ``bulk''
normal ordering where the divergences from image charges are not subtracted.
Therefore, they do not coincide with the original operators on the boundary,
even in the limit of $\ep\rightarrow 0$.
In the limit of $g(\theta)\rightarrow 0$, the relation between these two
different normal orderings
should be given by
\bea
O[X(\theta)]=\lim_{\ep\to
0}\exp{\left(\frac{1}{2}\log|2\ep|^2\left(\frac{\del}{\del X} \right)^2
\right)}:O[X(\theta,-\ep)]:.
\eea
The left-hand operator is defined by the boundary normal
ordering where effects of image charges are taken into
account.
Therefore, for the interaction terms of our model, we should multiply the
regularized operators (\ref{regint}) by a divergent factor, $1/\epp$.
This regularization is essentially the same one as in the original paper
\cite{Callan:1994ub}.

The boundary state $\bra{B}$ is defined by acting the exponential of the
boundary interaction (\ref{interaction})
on the Neumann boundary state $\bra{N}$,
\bea
\bra{B}=\bra{N}\exp\left(\half\int
d\theta\left[\frac{g(\theta)}{\epp}\exp{\left(\frac{iX(\theta,-\ep)}{\sqrt{2}}\right)}+\frac{\bar g(\theta)}{\epp}\exp{\left(\frac{-iX(\theta,-\ep)}{\sqrt{2}}\right)}\right]
\right).\label{defb}
\eea
Following the argument in \cite{Callan:1994ub},
we compactify the target space $X$ into the self-dual radius, $X\sim
X+2\pi\sqrt{2}$, and
perturbatively simplify this state using the $SU(2)$ current algebra.
In the $O(g(\theta))$ term, we have
\begin{align}
 & \half\bra{N}\int d\theta
\frac{g(\theta)}{\epp}:e^{i\frac{X(\theta,-\ep)}{\sqrt{2}}}:\nn \\
&= \half\bra{N}\int d\theta
\frac{g(\theta)}{\epp}:e^{i\frac{X_R(\theta,-\ep)}{\sqrt{2}}}::e^{i\frac{X_L(\theta,-\ep)}{\sqrt{2}}}:\nn \\
&= \half\bra{N}\int d\theta e^{-\frac{\pi
i}{4}}\frac{g(\theta)}{\epp}:e^{i\frac{X_L(\theta,\ep)}{\sqrt{2}}}::e^{i\frac{X_L(\theta,-\ep)}{\sqrt{2}}}:\nn \\
&=-i\pi\bra{N}\oint \frac{dz}{2\pi i}
g(z):e^{i\sqrt{2}X_L(z)}:.\label{deriveOg1}
\end{align}
In the second line of (\ref{deriveOg1}), we have separated the operator into
holomorphic and anti-holomorphic parts.
The second equality comes from a property of the Neumann boundary state, namely
$\bra{N}X_R(\theta,-\tau)=\bra{N}X_L(\theta,\tau)$.
The phase $e^{\frac{\pi i}{4}}$ is due to an effect of the normal
ordering\footnote{There is an ambiguity related to the choice of branches.
Here, we just pick up the one for reality of the boundary state.}
and it cancels another phase from the OPE among two holomorphic operators.
In the last line, we have used a holomorphic coordinate
$z=e^{-i\sigma+\tau}$, which is nothing but $e^{-i\theta}$ on the boundary.
We have also rewritten the external field in terms of $z$.
The full expression for the $O(g(\theta),\bar g(\theta))$ terms is given by,
\bea
-i\pi\bra{N}\oint \frac{dz}{2\pi i} \left(g(z)j^+(z)+\bar g(z)j^-(z)\right),
\eea
where $j^+(z)=:e^{i\sqrt{2}X_L(z)}:$ and $j^-(z)=:e^{-i\sqrt{2}X_L(z)}:$.
Note that this final expression does not contain divergent quantities any more.

Next, we shall consider the second-order terms,
\bea
\bra{N}\frac{1}{2!}S_{\rm{int}}S_{\rm{int}}
=-\frac{i\pi}{2!}
\bra{N}\oint \frac{dz}{2\pi i} \left(g(z)j^+(z)+\bar g(z)j^-(z)\right)
\half\int d\theta
\left[\frac{g(\theta)}{\epp}
:\exp{\left(\frac{iX(\theta,-\ep)}{\sqrt{2}}\right)}:
+{\rm c.c.}\right]
{}.
\eea
Because anti-holomorphic operators commute with holomorphic charges,
we can repeat the same argument as in (\ref{deriveOg1}) for these terms.
For the first half of these terms, we have
\bea
-\frac{i\pi}{2!}\bra{N}\frac{1}{2}\int d\theta \frac{e^{-\frac{\pi
i}{4}}}{\epp}\Bigg[g(\theta)\oint \frac{dz}{2\pi i} \left(g(z)j^+(z)+\bar
g(z)j^-(z)\right):e^{i\frac{X_L(\theta,\ep)}{\sqrt{2}}}::e^{i\frac{X_L(\theta,-\ep)}{\sqrt{2}}}:\nn \\
-g(\theta)\bar
g(\theta):e^{-i\frac{X_L(\theta,\ep)}{\sqrt{2}}}::e^{i\frac{X_L(\theta,-\ep)}{\sqrt{2}}}:\Bigg].
\eea
The main difference from the previous calculations is that,
we have an OPE between operators with the opposite charges.
This kind of OPE brings some divergent terms as follows,
\bea
\frac{(i\pi)^2}{2!}\bra{N}\oint \frac{dw}{2\pi i} \left[g(w)j^+(w)\oint
\frac{dz}{2\pi i} \left(g(z)j^+(z)+\bar g(z)j^-(z)\right)
-g(w)\bar g(w)(\frac{1}{2\ep w}+j^3(w))\right],
\label{evasec}
\eea
where $j^3(w)$ is one of the $SU(2)$ currents defined by $i\del X(w)/\sqrt{2}$.
The whole expression for the second-order terms is given by,
\bea
\bra{N}\frac{1}{2!}\left[\left(-i\pi\oint \frac{dz}{2\pi i}
\left(g(z)j^+(z)+\bar g(z)j^-(z)\right)\right)^2-2(i\pi)^2(2\ep)^{-1}\int
\frac{d\theta}{2\pi} g(\theta)\bar g(\theta)
\right].
\eea
While the linear term of $j^3(w)$ in (\ref{evasec}) has been canceled out by
its complex conjugate,
we still have a divergent term.

In principle, we can continue these arguments and obtain, order by order, the
expression of the boundary state, which is represented by $SU(2)$ charges.
However, as the order becomes higher, the expression becomes more complex and
the divergences become worse.
For example, for the third order terms, we have
\bea
\bra{N}\frac{1}{3!}\Bigg[\left(-i\pi\oint \frac{dz}{2\pi i}
\left(g(z)j^+(z)+\bar g(z)j^-(z)\right)\right)^3
+(-i\pi)^3\oint \frac{dz}{2\pi
i}\left(gg\bar{g}(z)j^+(z)+\bar{g}\bar{g}g(z)j^-(z)\right)\nn\\
-6(-i\pi)^3(2\ep)^{-1}\int\frac{d\theta}{2\pi} g(\theta)\bar g(\theta)
\oint \frac{dz}{2\pi i} \left(g(z)j^+(z)+\bar g(z)j^-(z)\right)
\Bigg]
{}.
\eea

In the next section, we shall derive the whole expression of this boundary
state to all orders in perturbation theory
using some differential equations
and explicitly show that we can remove the divergences by introducing only one
counter term corresponding to a constant shift of the boundary potential.

\section{Boundary State to All Orders in Perturbation Theory}

In the original work \cite{Callan:1994ub} on the conformal invariant
theory (\ref{originalBSG}), it is proved
that the above divergences are absorbed as a constant shift in the potential,
and terms in the perturbation series
are exponentiated and take the following simple form,
\bea
\bra{B}=\bra{N}e^{-i\pi (g_rJ_0^++\bar{g}_rJ_0^-)},\label{oldexpression}
\eea
where $g_r$ and $\bar{g}_r$ are renormalized couplings which should be
non-trivial functions of original couplings $g$ and $\bar g$.
However, in their work, the explicit dependence of the constant shift and the
renormalized couplings on the original couplings has not been given.
Though the constant shift of the potential does not affect the physics of this
two dimensional theory,
it should correspond to a condensation of the spatially homogeneous tachyon in
the string-theory's point of view.
Thus, it seem to be worth while to investigate it in more detail.
Furthermore, in this paper, we consider the generalized theory with the
coordinate-dependent interaction (\ref{interaction}).
Then, the relation between ``couplings'' becomes more important,
because  ``couplings'' are not just constant but some functions of $\theta$,
and the relation can contain more information.

In order to obtain the expression for the boundary state of our generalized
model to all orders of original couplings $g(\theta)$ and $\bar g(\theta)$,
we shall introduce a new technique here.
The previous result (\ref{oldexpression}) and our perturbative results in the
previous section suggest that the boundary state in terms of original couplings
is given by
\bea
\bra{B}=\bra{N}\exp\left[\int
d\theta\left(\half\ap(g(\theta),\bar{g}(\theta))e^{i\sqrt{2}X_L(\theta)}
+\half\bar{\ap}(g(\theta),\bar{g}(\theta))e^{-i\sqrt{2}X_L(\theta)}+\bt(g(\theta),\bar{g}(\theta))\right)\right].\label{ansatz}
\eea
The functions, $\ap(\theta)$ and $\bar \ap(\theta)$, are some functions which
are fixed by the original external fields,
$g(\theta)$ and $\bar g(\theta)$.
They are reduced to the
renormalized couplings, $g_r$ and $\bar{g}_r$, when external fields
are just constants.
We have added a function, $\bt(g,\bar{g})$, in order to represent possible
constant shifts of the potential.
Our strategy is to find differential equations which our functions, $\ap(g,\bar
g)$, $\bar \ap(g,\bar g)$ and $\bt(g,\bar g)$, should satisfy by comparing the
two expressions of the derivative of the boundary state.
Using these differential equations, we can obtain the exact expression for the
boundary state.

First, we start from the original definition (\ref{defb}) of the boundary
state.
By differentiating it with respect to the external field, $g(\theta)$, we have
\bea
\frac{\delta}{\delta g(\theta)}\bra{B}
=\bra{B}\frac{1}{2\epp}\int_0^1ds
e^{\frac{-s}{2}\int
d\theta\left(\frac{g}{\epp}e^{\frac{iX}{\sqrt{2}}}+\frac{\bar{g}}{\epp}e^{\frac{-iX}{\sqrt{2}}}\right)}
e^{\frac{iX(\theta)}{\sqrt{2}}}
e^{\frac{s}{2}\int
d\theta\left(\frac{g}{\epp}e^{\frac{iX}{\sqrt{2}}}+\frac{\bar{g}}{\epp}e^{\frac{-iX}{\sqrt{2}}}\right)}.
\eea
Especially, the following combination of the derivatives takes a simple form:
\bea
\int d\theta \left(g(\theta)\frac{\delta}{\delta g(\theta)}+\bar
g(\theta)\frac{\delta}{\delta \bar g(\theta)}\right)\bra{B}=
\bra{B}\frac{1}{2}\int
d\theta\left(\frac{g(\theta)}{\epp}e^{\frac{iX(\theta)}{\sqrt{2}}}+\frac{\bar{g}(\theta)}{\epp}e^{\frac{-iX(\theta)}{\sqrt{2}}}\right).
\eea
In the r.h.s of the above expression, we can replace the original definition
for $\bra{B}$ with the ansatz (\ref{ansatz}).
Because the holomorphic operators, $e^{\pm \frac{iX_L}{\sqrt{2}}}$, behave as
spin $1/2$ states for the holomorphic $SU(2)_L$ charges,
using the following equation,
\bea
\exp\left(-i\pi \Pmatrix{0&\ap\\ \bar\ap&0}\right)
=\Pmatrix{\cos(\pi\sqr)&-i\sqrt{{\ap}/{\bar\ap}}\sin(\pi\sqr)\\
-i\sqrt{{\bar\ap}/{\ap}}\sin(\pi\sqr)&\cos(\pi\sqr)},
\eea
we can easily repeat the same arguments made in the previous section.
Thus, the combination of the derivatives of the boundary state is given by
\bea
&&-i\pi\bra{B}\oint \frac{dz}{2\pi
i}\Bigg(g\cos(\pi\sqrt{\ap\bar\ap})j^+(z)+\bar
g\cos(\pi\sqrt{\ap\bar\ap})j^-(z)\nn\\
&&\phantom{aaaaaa}-i\sin(\pi\sqrt{\ap\bar\ap})
\left(\frac{1}{2\ep z}\left(g\sqrt{\bar\ap/\ap}+\bar g\sqrt{\ap/\bar
\ap}\right)
+(g\sqrt{\bar\ap/\ap}-\bar g\sqrt{\ap/\bar
\ap})j^3(z)\right)\Bigg).\label{deforiginal}
\eea
We assume that the two functions $\ap(g,\bar g)$ and $\bar\ap(g,\bar g)$ are
related by a single function $f(x)$ which depends only on the absolute value of
$g(\theta)$, $x(\theta)=|g(\theta)|$, namely,
\bea
\ap(g,\bar g)=gf(x), \qquad\bar \ap(g,\bar g)=\bar gf(x),\label{apsimple}
\eea
so that the last term of (\ref{deforiginal}) vanishes.
Next, we shall evaluate the same combination of the derivatives of the boundary
state
by differentiating the ansatz (\ref{ansatz}) directly.
The above assumption (\ref{apsimple}) about the dependence of the functions
$\ap$ and $\bar \ap$ also makes this evaluation quite simple.
The quantity at issue is evaluated as follows:
\begin{align}
&-i\pi\bra{B}\int_0^1 ds e^{i\pi s\oint \frac{dz}{2\pi i} \left(\ap
j^++\bar{\ap}j^-\right)}
\oint \frac{dz}{2\pi i} \left(g\frac{d(xf)}{dx}j^++\bar
g\frac{d(xf)}{dx}j^-+\frac{2i}{z}x\frac{d \bt}{d x}\right)
e^{-i\pi s\oint \frac{dz}{2\pi i} \left(\ap j^++\bar{\ap}j^-\right)}\nn\\
&=-i\pi\bra{B}\oint \frac{dz}{2\pi i}\left(
g\frac{d(xf)}{dx}j^+(z)+\bar g\frac{d(xf)}{dx}j^-(z)
+\frac{2i}{z}x\frac{d \bt}{d x}
+i\pi\frac{\del x}{\del z}x\left(\frac{d(xf)}{dx}\right)^2\right).
\end{align}
Here, we have also assumed that the function $\beta(g,\bar g)$ depends only on
$x(\theta)$.
Comparing these two expressions for the derivative of the boundary state, we
have following differential equations
\begin{align}
\frac{d(xf(x))}{dx}&=\cos(\pi xf(x)), \label{defeqap}\\
\frac{d \bt(x)}{d x}&=\frac{1}{2\ep}\sin(\pi xf(x))
-\frac{\pi z}{2}\frac{\del x}{\del z}\left(\frac{d(xf)}{dx}\right)^2
.\label{defeqbt}
\end{align}

Using these two differential equations, we can fix the functions,
$\ap(\theta)$, $\bar \ap(\theta)$ and $\bt(\theta)$,
in terms of the external fields, $g(\theta)$ and $\bar g(\theta)$.
With the suitable initial condition, the solution of (\ref{defeqap}) is given
by
\bea
f(x)=\frac{2}{\pi x}\arctan\left[\tanh\left(\frac{\pi}{2}x\right)\right].
\eea
Thus, we can determine the functions, $\ap(\theta)$ and $\bar\ap(\theta)$,
which appear in the exponent of the boundary state, as
\begin{align}
\ap(g(z),\bar
g(z))=&\frac{2}{\pi}\frac{g(z)}{|g(z)|}\arctan\left[\tanh(\frac{\pi}{2}
|g(z)|)\right],\\
\bar\ap(g(z),\bar g(z))=&\frac{2}{\pi}\frac{\bar
g(z)}{|g(z)|}\arctan\left[\tanh(\frac{\pi}{2} |g(z)|)\right].
\end{align}
This relation is one of our main results in this paper.
If we turn off the non-zero modes of the external fields, $g(z)$ and
$\bar{g}(z)$, these equations give the relation between
the renormalized couplings, $g_r$ and $\bar{g}_r$, and the bare ones,
$g$ and $\bar g$:
\begin{align}
g_r=&\frac{2g}{\pi|g|}\arctan\left[\tanh(\frac{\pi}{2}
|g|)\right],\label{ggr-relation1}\\
\bar{g}_r=&\frac{2\bar g}{\pi|g|}\arctan\left[\tanh(\frac{\pi}{2}
|g|)\right].\label{ggr-relation2}
\end{align}
In the small coupling limit, $|g|\rightarrow 0$, $g_r$ approaches $g$.
On the contrary, in the large coupling limit, $|g|\rightarrow \infty$, the
renormalized coupling,
$|g_r|$, approaches $1/2$.
This property is exactly what the authors of \cite{Callan:1994ub} has
conjectured for the renormalized couplings
as a finite renormalization effect.
They have shown that the boundary state (\ref{oldexpression}) represents the
Dirichlet boundary condition
when its renormalized coupling $|g_r|$ is $1/2$.
On the other hand, since, in the large coupling limit,
the interaction potential forces the boundary value of the field to be fixed,
the corresponding boundary state should be the Dirichlet boundary state.
It means that the renormalized coupling should approach $1/2$ in this limit.
Our results (\ref{ggr-relation1}) and (\ref{ggr-relation2}) explicitly describe
this renormalization effect.

Next, we consider the equation (\ref{defeqbt}) for constant shifts of the
boundary potential.
It is convenient to split the function $\beta$ as $\beta=\beta_1+\beta_2$, and
to consider the following two equations:
\begin{align}
\frac{d \bt_1(x)}{d x}=&\frac{1}{2\ep}\sin(\pi xf(x)),\\
\frac{d \bt_2(x)}{d x}=&-\frac{\pi z}{2}\frac{\del x}{\del
z}\left(\frac{d(xf)}{dx}\right)^2=-\frac{z}{2}\frac{\del}{\del z}\sin(\pi
xf(x)).
\end{align}
The integrations of them give
\begin{align}
\beta_1(g(z),\bar g(z))=&\frac{1}{2\pi\ep}\log(\cosh(\pi |g(z)|)),\\
\beta_2(g(z),\bar g(z))=&-\frac{z}{2\pi}\frac{\del}{\del z}\log(\cosh(\pi
|g(z)|)).
\end{align}
The function $\bt_1$ depends on the regularization parameter $\ep$ and
diverges in the limit of $\ep\rightarrow 0$.
However, this divergence is absorbed by a counter-term, or a constant shift of
the boundary interaction
and it is harmless, as expected.
Furthermore, the function $\bt_2$ affects nothing, because it gives a
integration of a total derivative.
{}From the view point of string theory, the function $\bt$ corresponds to the
spatially homogeneous tachyon.
It tells us how much we should add the spatially homogeneous tachyon in
addition to the space-dependent one.
Equivalently, this information can be absorbed into a definition of the
boundary operators.
For example,
\bea
e^{iX(\theta)/\sqrt{2}}\equiv\lim_{\ep\rightarrow 0}\left[
\frac{1}{\sqrt{2\ep}}:e^{iX(\theta,-\ep)/\sqrt{2}}:
-\frac{1}{2\pi\ep g(\theta)}\log(\cosh(\pi |g(\theta)|))
\right], \label{defofbo}
\eea
and a similar definition for the operator $e^{-iX(\theta)/\sqrt{2}}$.
Under such a prescription,
our final expression for the boundary state for the theory with the interaction
(\ref{interaction}) is
\begin{align}
\bra{B}=&\bra{N}\exp\Bigg[-2i\oint \frac{dz}{2\pi i}
\Bigg(\frac{g(z)}{|g(z)|}
\arctan\left(\tanh(\frac{\pi}{2} |g(z)|)\right)j^+(z)\nn\\
&\hspace{4cm}+\frac{\bar g(z)}{|g(z)|}\arctan\left(\tanh(\frac{\pi}{2}
|g(z)|)\right)j^-(z)\Bigg)\Bigg].
\label{expforbs}
\end{align}
Strictly speaking, our procedure is not strong enough to determine this
boundary state completely, except for the case that $g(\theta)$ is just real
constant $g$.
Our differential equations are nothing but consistency conditions on our ansatz
(\ref{ansatz}).
However, we have checked that our result (\ref{expforbs}) correctly reproduces
the perturbative results in the previous section to at least first few orders.
Therefore, we just conjecture that our result (\ref{expforbs}) is the correct
one even with external fields, $g(\theta)$ and $\bar{g}(\theta)$.

In fact, we can directly check whether the state (\ref{expforbs}) satisfies the
boundary condition derived from the action with external fields.
The boundary condition is given by
\bea
\frac{1}{\sqrt{2}\pi}\frac{dX(\theta)}{d\tau}-ig(\theta)e^{i\frac{X(\theta)}{\sqrt{2}}}+i\bar g(\theta)e^{-i\frac{X(\theta)}{\sqrt{2}}}=0.\label{bc}
\eea
However, there is an ambiguity related to the definition of the boundary
operators.
Because this boundary condition is for the interacting theory, the boundary
operators appearing in the above condition
must be the operators which are defined in this interacting theory.
We prefer to define these operators using differentiation with respect to the
external fields.
Under this prescription, the boundary condition for the boundary state is
represented as
\bea
\bra{B}\frac{-1}{\sqrt{2}\pi}\frac{dX(\theta)}{d\tau}=2\left(ig(\theta)\frac{\delta}{\delta g(\theta)}-i\bar g(\theta)\frac{\delta}{\delta \bar g(\theta)}\right)\bra{B}.\label{bcforbs}
\eea
A short calculation leads the both sides of (\ref{bcforbs}) equally into the
following form:
\begin{align}
&-\frac{z}{\pi}\bra{B}\Bigg(
\half\sqrt{\frac{\ap}{\bar \ap}}\sin(2\pi\sqrt{\ap\bar\ap})j^+(z)
-\half\sqrt{\frac{\bar \ap}{\ap}}\sin(2\pi\sqrt{\ap\bar\ap})j^-(z)
\nn\\
&\hspace{1cm}-2i\sin^2(\pi\sqrt{\ap\bar\ap})j^3(z)
+\frac{i}{2}\sin^2(\pi\sqrt{\ap\bar\ap})\left(\frac{\del\bar\ap}{\bar\ap}-\frac{\del\ap}{\ap}\right)\Bigg).
\end{align}
Thus, our ``boundary state'' actually satisfies the boundary condition.
Note that, in this calculation, we have no need to use the concrete relations
(\ref{ggr-relation1}) and (\ref{ggr-relation2}) between external fields and
renormalized ones.
Therefore, we cannot fix their relations from this condition.

\section{Disc Partition Function and Correlation Functions}
In this section, we shall consider the disc partition function and some
correlation functions\footnote{
More general correlation functions of the original model (\ref{originalBSG})
has been discussed in \cite{Callan:1994ub,Kristjansson:2004mf}.
} of this model.
To calculate the disc partition function $Z[g(\theta)]=\bracket{B}{0}$, we must
reexpress the boundary state only in terms of non-negative modes.
In the case that $g(\theta)$ is a real function, we can easily do it using our
expression (\ref{expforbs}).
In the rest of this section, we shall focus on such a case.
Then, the boundary state can be rewritten as
\begin{align}
\bra{B}=&\bra{N}\exp\left(-2\pi i\sum_{n=-\infty}^{\infty}\ap_n
J^1_n\right)\nn\\
=&\bra{N}\exp\left(2\pi i\sum_{n=1}^\infty\ap_{-n}\tilde{J}^1_n\right)
\exp\left(-2\pi i\sum_{n=1}^\infty \ap_nJ^1_n\right)
\exp\left(i\pi\ap_0\left(\tilde{J}^1_0-{J}^1_0\right)-4\pi^2\sum_{n=1}^\infty
n\ap_n\ap_{-n}\right),
\end{align}
where $J^1_n$($\tilde{J}^1_n$) are charges of
holomorphic(anti-holomorphic) $SU(2)$ symmetry and the coefficients
$\ap_n$ are Fourier coefficients of the function $\ap(\theta)$, namely,
$\ap(\theta)=\sum_n \ap_n e^{-in\theta}$. Using the free energy,
\bea
w[g(\theta)]=4\pi^2\sum_{n=1}^\infty n\ap_n\ap_{-n}
=4\pi^2\sum_{n=1}^\infty n\int \frac{d\theta}{2\pi}\int
\frac{d\theta'}{2\pi}\ap(\theta)\ap(\theta') \cos{n(\theta'-\theta)},
\eea
the disc partition function of this system is written as
\bea
Z[g(\theta)]_{\rm Disc}=\bracket{B}{0}=\exp\left(-w[g(\theta)]\right),
\label{partitionfunction}
\eea
where we have simply normalized the inner product $\bracket{N}{0}$, which is
independent of $g(z)$, as 1.
If the external field is just constant, the partition function (or the free
energy) does not depend on the coupling $g$.
In on-shell case, we can regard a partition function as an action for a string
field theory \cite{bsft}.
The independence of the partition function from the constant coupling $g$ just
tells us that, at the self-dual radius,
the total energy of the D-brane does not change during the tachyon condensation
into the lower dimensional D-brane \cite{Sen:1999mh}. Note that, for this
independence, we must carefully select the tachyon profile  (\ref{defofbo}).

The 1-point function of the boundary operator is directly calculated from the
derivative of the free energy:
\bea
\left\langle\cos\left(\frac{X(\theta)}{\sqrt{2}}\right)\right\rangle=-\frac{\delta w}{\delta g(\theta)}
=-4\pi^2 \cos(\pi\ap(\theta))\sum_{n=1}^\infty
n\int\frac{d\theta'}{2\pi}\ap(\theta') \cos{n(\theta'-\theta)}.
\eea
This 1-point function vanishes if we take the external field as a constant.
The connected 2-point function is given by
\begin{align}
\left\langle\cos\left(\frac{X(\theta_1)}{\sqrt{2}}\right)\cos\left(\frac{X(\theta_2)}{\sqrt{2}}\right)\right\rangle_{\rm connected}
=&4\pi^2\cos(\pi\ap(\theta_1))\cos(\pi\ap(\theta_2))\frac{1}{\sin^2(\frac{\theta_1-\theta_2}{2})}\nn\\
&+\pi\delta(\theta_1-\theta_2)\sin(\pi\alpha(\theta_1))\left\langle\cos\left(\frac{X(\theta_1)}{\sqrt{2}}\right)\right\rangle.
\label{2-point}
\end{align}
These correlation functions agree with the usual ones in the free limit of
$g(\theta)\rightarrow0$.
In this limit, 1-point function just vanishes and 2-point function
becomes\footnote{
Note that these arguments are rather formal.
We have freely used integrations by parts to derive the 2-point function
(\ref{2-point}).
We should evaluate well-defined quantities with integrations over the
coordinate.}
\bea
\left\langle\cos\left(\frac{X(\theta_1)}{\sqrt{2}}\right)\cos\left(\frac{X(\theta_2)}{\sqrt{2}}\right)\right\rangle_{\rm free}
\sim\frac{1}{\sin^2(\frac{\theta_1-\theta_2}{2})},
\eea
as expected.
On the other hand, in the large coupling limit of $g\rightarrow\infty$ of the
constant coupling,
they cease to propagate, where $\cos(\pi\ap)=0$.

Before ending this section, we shall report a curious property of the partition
function (\ref{partitionfunction}).
The Lagrangian for our model has an explicit dependence on the boundary
coordinate through the external field $g(\theta)$.
Therefore, this model looks like a model which does not have conformal
invariance except for the case that $g(\theta)$ is a constant.
However, our disc partition function still knows the conformal invariance in
the following sense.
Let us consider the $SL(2,R)$ transformations of the external field, which map
the boundary of the disk into itself.
The infinitesimal transformation of the external field is given by
\bea
\delta g(z) = (\ep_0+\ep_1 z+\ep_2 z^2)\del g(z),
\eea
where parameters satisfy
$\bar{\ep}_0=-\ep_2, \bar{\ep}_1=-\ep_1$.
It is easily shown that these transformations of the external field do not
change the partition function (\ref{partitionfunction}).
This fact suggests that, though our theory does not possesses the full symmetry
under the conformal transformation,
it still has a symmetry under the subgroup of the transformation,
namely, $SL(2,R)$.

\section{Summary and Discussion}
In this paper, we have extended the argument by \cite{Callan:1994ub} into the
case where their couplings are replaced with arbitrary periodic functions.
Even in this generalized theory, their argument to rewrite the state in terms
only of holomorphic fields still works.
Using differential equations, we have obtained the boundary state, to all
orders in perturbation theory,
with the relations between the ``bare external fields'' coupled
to the boundary interaction and the ``renormalized external fields'' which form
charges with $SU(2)$ currents.
This explicit determination of the boundary state for the generalized theory
(\ref{interaction}) is our main result.
Using this boundary state, we have also calculated the disc partition
function (or the free energy) of this system.

So far we have solved this theory, to all orders in perturbation theory, in
spite of the existence of the boundary interaction (\ref{interaction}).
However, this is not so surprising.
The free energy $w[g(\theta)]$ of this system is written by a quadratic form of
$\ap(\theta)$.
This fact suggests that there exists a free theory coupled to the external
field, not $g(\theta)$, but $\ap(\theta)$,
which becomes equivalent to our theory after path integration.
Actually, such a theory exists for the original theory (\ref{originalBSG}).
Using fermionization, the authors of \cite{Polchinski:1994my} mapped the theory
into a fermionic system with quadratic interactions.
Our results in this paper give a precise relation between these two theories.

In the presence of external fields, the corresponding boundary
state does not satisfy re\-parametriza\-tion invariance conditions,
$\bra{B}(L_n-\tilde{L}_{-n})=0$. Thus, this state should be
an off-shell state, though it satisfies the boundary condition (\ref{bc}).
Although, at present, we have no specific applications of this model to the
field of string theory, it would be interesting to find such
candidates.
Besides, in the original paper \cite{Callan:1994ub}, the theory with conformal
symmetry was studied in detail as a quantum field theory of two dimensional
space-time. Similar kinds of analyses would be applicable to our generalized
one.

Or rather, our generalized theory might be a suitable tool for investigations
into the original one with conformal invariance.
For example, we can simply define the local boundary operators of the
interaction theory from the derivatives with respect to corresponding external
fields without difficulties of divergences.
And, there, all quantities are written in terms of ones appearing in the action
(\ref{originalBSG}).
Though these results are rather technical, they might help us
to study more important physics such as rolling tachyon.

\section*{Acknowledgments}
We would like to thank M.\ Fujita, M.\ Fukuma, H.\ Hata, P.\ Ho and S.\
Sugimoto for valuable discussions and
comments.
S.\,T. also thanks T.\ Noguchi, J.\ Wang and T.\ Yamashita.
The work of S.\,T. was supported by the National Center for Theoretical
Sciences at Taipei.

\end{document}